\documentclass[12pt, draftcls, onecolumn]{IEEEtran}
\usepackage{graphicx} 
\usepackage{epsfig} 
\usepackage{mathptmx} 
\usepackage{times} 
\usepackage{amsmath} 
\usepackage{amssymb}  
\usepackage{color}
\usepackage{authblk}
\usepackage[caption=false,font=footnotesize]{subfig}
\usepackage{setspace}
\usepackage{algorithm}
\usepackage{algorithmic}
\DeclareMathOperator*{\argmax}{argmax}

\newtheorem{theorem}{Theorem}
\newtheorem{lemma}{Lemma}

\newcommand{\field}[1]{\ensuremath{\mathbb{#1}}}
\newcommand{\N}{\ensuremath{\field{N}}} 
\newcommand{\E}[1]{{\mathbb{E}}\left[{#1}\right]}

\newcommand{\Prob}[1]{{\mathbb{P}}\left({#1}\right)}

\begin{document}
\doublespacing
\title{{ Self-sufficient Receiver with Wireless Energy Transfer in a Multi-access Network}}

\author{{ Yunus Sarikaya and Ozgur Ercetin}%
\thanks{Y. Sarikaya (yunus.sarikaya@unimelb.edu.au) is with the Department of Electrical and Electronic Engineering, University of Melbourne, Australia}  \thanks{O. Ercetin (oercetin@sabanciuniv.edu)  is with the Department of Electronics Engineering, Sabanci University, Turkey.} }%

\maketitle

\vspace{-2cm}
\begin{abstract}
In this letter, we consider the control of an energy self-sufficient receiver in a multi-access network with simultaneous wireless information and energy transfer.  Multiple transmitters send data to a common receiver whose only source of energy is a finite size battery which is recharged only from the energy harvested from incoming RF signals.  The nodes access the channel randomly resulting in a packet collision when multiple transmitters simultaneously access the channel. The receiver takes samples from the received RF signal to calculate the probability of a collision.  The objective is to maximize the receiver goodput subject to the instantaneous availability of receiver energy. We develop an asymptotically optimal dynamic control algorithm, where the receiver makes an energy harvesting or decoding decision according to the current channel measurements and battery level.



\end{abstract}
{\allowdisplaybreaks\vspace{-0.1in}
\section{Introduction}

Energy harvesting (EH) communications offer the promise
of energy self-sustaining operation
for wireless networks with significantly prolonged lifetimes.
There has been a plethora of work developing algorithms for EH communication networks. Most Prior works considered an {\em offline} control framework for EH transmitters, wherein the exact or statistical characteristics of the EH processes were assumed to be known a priori, e.g., \cite{Lei09,Sharma10}.
In \cite{Wang12, Aprem13}, the
statistics governing the random processes are assumed to be
available at the transmitter, while their realizations are only known
causally. An EH communication system was usually modeled as a Markov decision process (MDP), and a dynamic programming (DP) formulation was used to optimize the throughput numerically. However, the complexity of the employed solution is usually prohibitively high due to dimensionality curse of DP.  The authors in \cite{Huang13, Mao12} develop low-complexity algorithms for energy-harvesting networks by relying on Lyapunov optimization techniques, and show that their algorithms achieve near-optimal solution. 

The emphasis
on minimizing transmission energy is reasonable in traditional
wireless networks where the transmission distance is large, so that the transmission energy is dominant in
the total energy consumption. However, in many recently emerging technologies such as multi-tier heterogeneous networks and micro sensor networks,
the nodes are densely distributed with short-distances between the nodes. In such systems, the
circuit energy consumption due to encoding/decoding processes becomes
comparable to or even dominates the transmission energy
in the total energy consumption \cite{Cui_04}. There are a few number of studies investigating the operation of EH receivers, e.g., \cite{Tutuncuoglu12} and \cite{Arafa15}. However, these focus on the optimization of the operation over a single link with one transmitter and one EH receiver.


Most prior works analyze and design an EH system by assuming that the energy and information sources are uncorrelated, which is mainly true if the energy source is a natural source. In this letter, we jointly consider the wireless energy and information transfer in a { \em multi-access} network over a random access channel. Each node transmits with a certain probability to a common EH receiver with multiple simultaneous transmissions results in a collision and complete loss of information from all transmissions \footnote{This is a practically relevant model representing many of the common wireless networks such as omnipresent IEEE 802.11 based wireless networks.}.  The receiver harvests energy from incoming transmissions to recharge its finite size battery or attempts to decode the incoming message by consuming a portion of energy stored in its battery.  
Intuitively, if the receiver has full knowledge of the channel conditions, it would perform energy harvesting in those time-slots when there is a collision, since the received signal cannot be decoded anyway. However, in general, the receiver is not aware of a collision before attempting to decode it first. In which case, the received signal can no longer be harvested for energy.  

In order to alleviate this problem, we consider the case when the receiver takes samples from the channel to measure the received power at the beginning of each time-slot in order to calculate the probability of a collision as a function of the measured received power. Our objective is to maximize the average throughput, i.e., the average rate of packets which are successfully decoded by considering the availability of energy in the receiver battery, and the received RF signal power. By following similar approaches as in \cite{Huang13, Mao12}, and integrating the Lyapunov optimization techniques, we develop a stochastic online control scheme for multi-access channels with energy-harvesting receiver. To the best of our knowledge,
our work is the first provably asymptotically optimal scheme that handles energy-harvesting receiver in a multi-access network. Rigorously, we establish the performance guarantees of the proposed scheme, and show that it achieves a throughput that is within $O(\frac{1}{V})$ of the optimal, for any $V>0$.  

}
{\allowdisplaybreaks\vspace{-0.5cm}
\section{System Model}
\label{sec:model}

We consider a wireless uplink network, where $N$ transmitters aim to send their information to a common receiver in a distributed fashion. Furthermore, the transmitters are in close proximity of each other so that they can reach the receiver in a single hop, but they also cause excessive interference to each other when two or more transmitters are active at the same time. This leads to a fully connected interference graph topology with collision model for the wireless network in question. The receiver is capable of harvesting energy from the wireless transmissions, and storing it in a finite capacity rechargeable battery. It has no additional
power supplies. Furthermore, the receiver spends energy while attempting to decode and store the incoming message.

Time is slotted with slot index represented by $t \in \N$. The link quality between the transmitters and the receiver varies over time according to the {\em block} fading model, in which the channel gain is constant over a time slot and changes from one slot to another independently according to Rayleigh fading distribution. We use $h_i(t)$, $i = 1, \ldots, N$, to represent the gain of the channel between the transmitter $i$ and the receiver. The channel gains are exponentially distributed with mean $\mu_i$ for transmitter $i = 1,\ldots, N$, and they are independent and identically distributed (iid) over time. The transmit powers are  constant, identical to $P$ over all slots $t$. We normalize the power gains such that the (additive Gaussian) noise has unit variance. 
Then, the achievable rate of transmitter $i$ in time slot $t$, $R_i(t)$ is equal to $\log(1+Ph_i(t))$ bit$/$channel use\footnote{Our results can be applied for other channel models with minor modifications.}.  We assume that each transmitter has perfect knowledge of its channel gain, but not of others. Transmitters may accurately estimate their channel to the receiver by using the pilot-aided technique \cite{Morelli09}. Let $\phi_{pi}$ be the energy consumed for pilot signal transmission by the receiver. Note that only the distributions of channel gains are available at the receiver.

We consider a random access channel, where the transmitters access the channel  with a probability of $q_i$ for all $i$ at every slot $t$
\footnote{Here, more intelligent ways of transmission can be considered such as threshold policies as in \cite{Qin06}. We leave the analysis of other possible transmission policies as a future work.}.  Thus, let $S(t) \in \{ 0,1 \}$ be the random variable denoting the  outcome of transmission at $t$, with $S(t) = 1$ if the transmission is successful and $S(t) = 0$ otherwise \footnote{There is a constant drain from the battery of the receiver due to pilot signal transmissions, so the receiver may not have enough energy for pilot transmissions. Hence, we assume that even though a pilot transmission is not transmitted in a given time, the transmitters will continue their usual transmissions.}.



At the beginning of each time-slot $t$, the receiver spends $\tau$ portion of the time-slot to sense the channel and estimate received  signal power. The number of samples collected to estimate the received power is  $\frac{\tau}{f_s}$, where $f_s$ is the sampling frequency. We assume that the number of samples collected is sufficiently high that the estimation error is negligibly small\footnote{Note that $\tau$ is a system parameter and is determined according to the requirement of sensing accuracy.}. Let $\phi_{se}$ be the energy consumed for sensing the channel, which is the same for all $t$. Let $\gamma(t)$ denote the estimated received power at $t$.  Furthermore, the receiver spends $\phi_{de}(\gamma(t))$ amount of energy to decode the incoming transmission, where $\phi_{de}$ is generally an increasing convex function of the rate, i.e., $\log(1+\gamma(t))$ \cite{Arafa15}. 
Based on the information about the received power, $\gamma(t)$, the receiver decides whether to use the remaining time-slot for energy-harvesting or decoding. Here, we assume that the receiver is equipped with an integrated receiver antenna \cite{Zhou13}, in which the information decoding and the energy harvesting circuits are integrated in a single antenna. Equipped with an integrated receiver antenna, the receiver can freely switch between the information and energy receptions. Thus, let $\rho(t)$ be an indicator variable representing the action of the receiver in slot $t$. Specifically, if $\rho(t) = 1$, the receiver decides to decode the received information, and  if $\rho(t) = 0$, the receiver decides to harvest energy from incoming RF transmission\footnote{Note that the receiver would never choose to remain idle, unless the channel is idle, since it should use every opportunity to recharge its battery.}.  

We model the finite capacity rechargeable battery of the receiver as an { \em energy queue}. Let $E(t)$ be the size of the energy queue, which is equal to the amount energy left in the battery in time-slot $t$. The energy queue evolves as follows:
\begin{align*}
E(t+1) = \left [E(t) - \rho(t)\phi_{de}(\gamma(t)) - \phi_{se} - \phi_{pi}  + (1-\rho(t))(1-\tau)\eta\gamma(t) \right]^+,
\end{align*}
where $[x]^+ = \max(0,x)$ and $0< \eta \leq 1$ is the efficiency of the EH circuitry.

}
{\allowdisplaybreaks\vspace{-0.15in}
\section{Maximizing the Throughput of EH Receiver}
\label{sec:control}

We aim at maximizing the long-term average total throughput
by taking into account the estimated received power and the available energy at the receiver. We aim to solve the following optimization problem:

\vspace{-0.25in}
\begin{align}
		\max_{\rho(t)}  & \ \ 		r = \lim_{T \rightarrow \infty} \frac{1}{T} \sum_{\tau = 1}^T \log(1 + \gamma(\tau))S(\tau)\rho(\tau) \label{eq:obj}\\
		\mbox{subject to } & \rho(t)\phi_{de}(\gamma(t)) + \phi_{se} + \phi_{pi} \leq E(t), \ \forall t \label{eq:energy_const},
\end{align}
where $\eqref{eq:energy_const}$ is the energy-availability constraint, i.e., the consumed energy must be no more than the available energy. The channels are ergodic and stationary, so the objective function in \eqref{eq:obj} can be rewritten as:
\begin{align*}
 r = \E{\log(1 + \gamma(t))\rho(t) P_s(t)},
\end{align*}
where $P_s(t)$ is the probability of successful decoding. 

\begin{lemma}
\label{lemma:prob}
		For given transmission probabilities, $q_1, \ldots, q_N$, and the received power, $\gamma(t)$, the successful decoding probability is:
		
		\vspace{-0.1in}
		\small
		\begin{equation}
		P_s(t) = \frac{\sum\limits_{i=1}^N \mu_i^{-1} e^{-\frac{\gamma(t)}{\mu_i}}q_i \prod\limits_{j \neq i} (1-q_j)}{\sum\limits_{k=1}^{2^N-1} \left[\sum\limits_{i\in A_k}\mu_i^{-1} e^{-\frac{\gamma(t)}{\mu_i}}\prod\limits_{j\in A_k, \\ j \neq i} \frac{\mu_i}{\mu_i - \mu_j}\right] \left[\prod\limits_{i \in A_k} q_i \prod\limits_{ n \notin A_k^m} (1-q_n)   \right] }, \nonumber
		\end{equation}
	\normalsize where $A_k$ is $k$th subset of active transmitters. Note that the total number of non-empty subsets is $2^N-1$.
\end{lemma}

 \begin{IEEEproof}  The proof of the lemma is given in Appendix \ref{proof:prob}. \end{IEEEproof}

Next, we present our optimal dynamic control policy for the solution of \eqref{eq:obj}-\eqref{eq:energy_const}. The proposed dynamic algorithm is based on the
stochastic network optimization framework \cite{Georgiadis}. The optimization problem can in principle be addressed by DP and$/$or Markov chain techniques. Note that the states representing the residual energy in the battery are continuous in our model. A continuous state MDP is a natural way to represent a decision
process over continuous space of states. However, generating
optimal policies tractably for such MDPs is not trivial. To address this challenge, suboptimal methods are proposed, which are based on  approximations on either states or transition state probabilities \cite{Toussaint06}. Furthermore, these methods are usually impractical, since they suffer from the dimensionality curse of DP resulting in extensive computational cost. On the other hand, stochastic network optimization framework allows the solution of a long-term stochastic optimization problem by maximizing each time-slot separately without considering the effect of the solution in later time-slots, and obtains an asymptotically optimal solution.

To this end, we first choose a {\em perturbation} variable, $\theta$ and define a quadratic perturbed Lyapunov
function as: $\frac{1}{2} \left(E(t) - \theta \right)^2$.
The intuition behind the use of $\theta$ is to push the size of energy queue, $E(t)$, towards $\theta$ by keeping the Lyapunov function value small. Thus, we can ensure that the energy queue always has energy for data reception by carefully choosing the value of $\theta$. Later, we will show that the size of the energy queue, $E(t)$, cannot be larger than $\theta$ plus some constant. Thus, it can be considered as the limiting variable for the maximum size of the energy queue.

Also, consider the one-step expected Lyapunov drift, $\Delta(t)$ for
the Lyapunov function as:
\vspace{-0.15in}
\begin{equation*} \Delta(t) =
\mathbb{E}\left[ L(t+1) -
L(t)|E(t)\right].
\end{equation*}
Here, the expectation is taken over the distribution of the channel state, as well as the randomness in choosing the control action.
The aim of stochastic optimization framework is to minimize the drift to ensure that there is always energy in the battery. This can be achieved by having a negative Lyapunov drift
whenever the difference between $E(t)$ and $\theta$ is sufficiently large. Furthermore, the following  throughput-mixed Lyapunov drift

\small
\vspace{-0.1in}
\begin{equation} \Delta^U(t)=\Delta(t) - V\E{(1-\tau)\log(1+\gamma(t))\rho(t)S(t) | \gamma(t),E(t)}, \label{eq:deltawithreward}
\end{equation}
\normalsize
enables us to maximize the network throughput in conjunction with the battery size. 

Next, we present the control algorithm that minimizes \eqref{eq:deltawithreward} and provide its optimality in Theorem \ref{thm:optimalcontrol}.

{\bf Control Algorithm:} The receiver observes the energy queue size, and the received power. Then, it determines its control decision, $\rho(t)$ as the solution of the following optimization problem:

\vspace{-0.3in}
\begin{align*}
\rho(t)= \argmax_{x \in \{ 0,1\}} \  & V (1-\tau)\log(1+\gamma(t))x P_s(t)   - (E(t)-\theta)\left( x\phi_{de}(\gamma(t)) -  (1-x)(1-\tau)\eta\gamma(t) \right) \},
\end{align*}
where $V > 0$ is a design constant that  will determine the final performance of the algorithm. The above problem is a simple index policy whose solution is given as:
\begin{align*}
	\rho(t) =\begin{cases} 1, & \text{ if  }   -(E(t)-\theta)\left((1-\tau)\eta\gamma(t) - \phi_{de}(\gamma(t))  \right) \leq   V (1-\tau)\log(1+\gamma(t))P_s(\gamma(t)) \\
0, & \text{otherwise}
\end{cases}
\end{align*}

 
Next, we present the performance bound of the proposed algorithm.

\begin{theorem}
\label{thm:optimalcontrol}
Under the proposed algorithm with $\theta = \frac{V}{\eta} + \phi_{de}(\gamma_{max}) + \phi_{se}+\phi_{pi}$  we have the following:
\begin{enumerate}
	\item[\bf (1)] The energy queue at the receiver satisfies 
	$0 \leq E(t) < \theta + \gamma_{max},$
where $\gamma_{max}$ is the maximum received power. Moreover, when $E(t) \leq \phi_{de}(\gamma_{max}) + \phi_{se}+\phi_{pi}$, the receiver always decides to harvest energy, 
\item[\bf (2)]  Suppose $\bar{r}$ is the average throughput achieved by the proposed dynamic control algorithm. Then, for any  $V > 0$,  the dynamic control algorithm yields the following performance bound:
\vspace{-0.1in}
 \begin{align*}
   \bar{r} &\geq r^* - \frac{B}{V}
		\end{align*}		
 where $B > 0$ are constants, and $r^*$ is the optimal solution of the problem in
\eqref{eq:obj}. 
\end{enumerate}
\end{theorem}

 \begin{IEEEproof}  The proof of the theorem is given in Appendix \ref{proof:optimalcontrol}. \end{IEEEproof}

Theorem  \ref{thm:optimalcontrol} shows that the proposed dynamic control gets arbitrarily close to the
optimal utility with sufficiently large $V$ at the expense of larger battery capacity.

}
{\allowdisplaybreaks\section{Numerical Results}
\vspace{-1cm}
\begin{figure*}[htp]
\centerline{ \subfloat[Performance with respect to $V$,]{\includegraphics[width=3.5in]{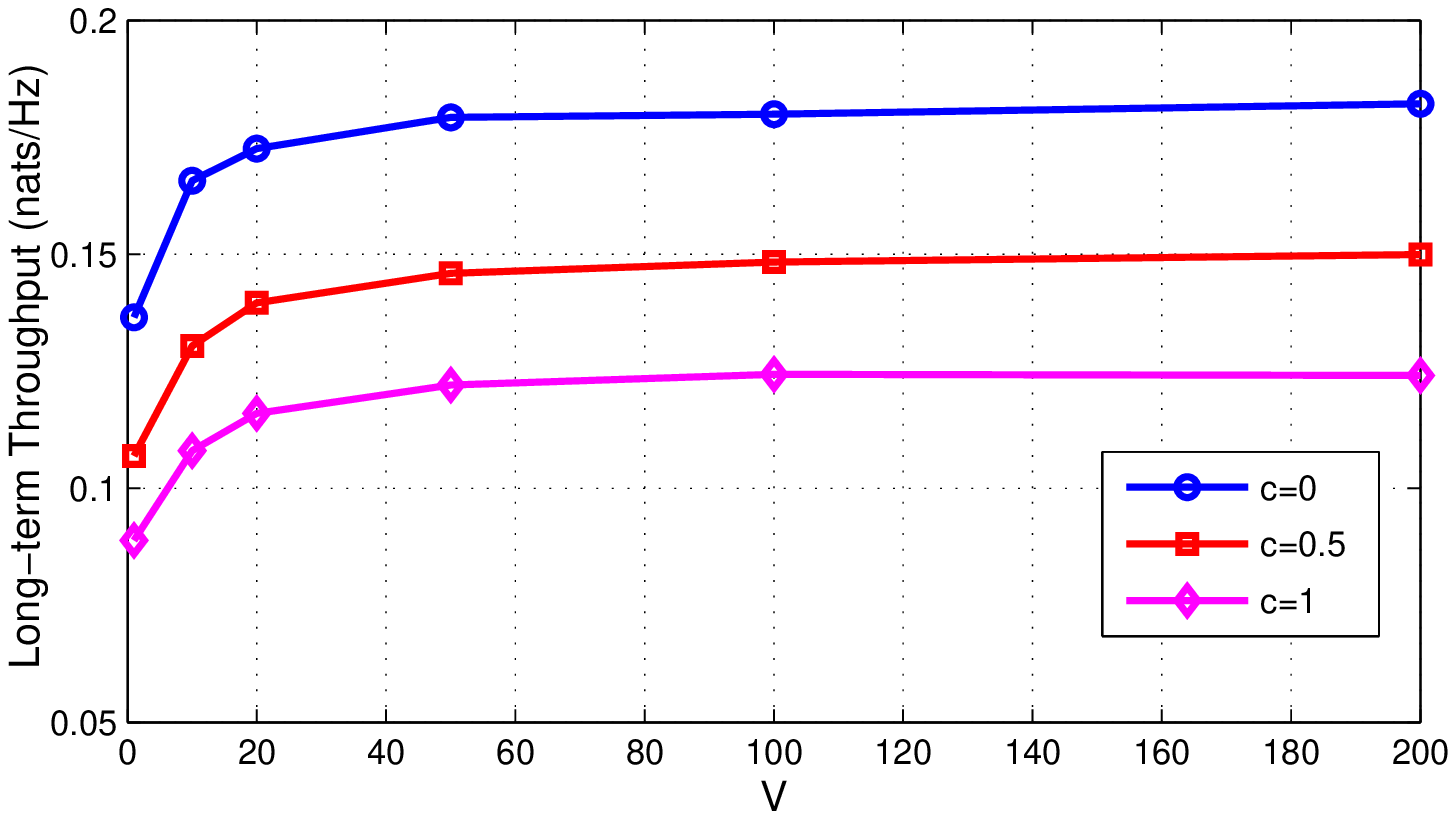}
\label{fig:V}}  
\subfloat[Performance with respect to $q$,]{\includegraphics[width=3.5in]{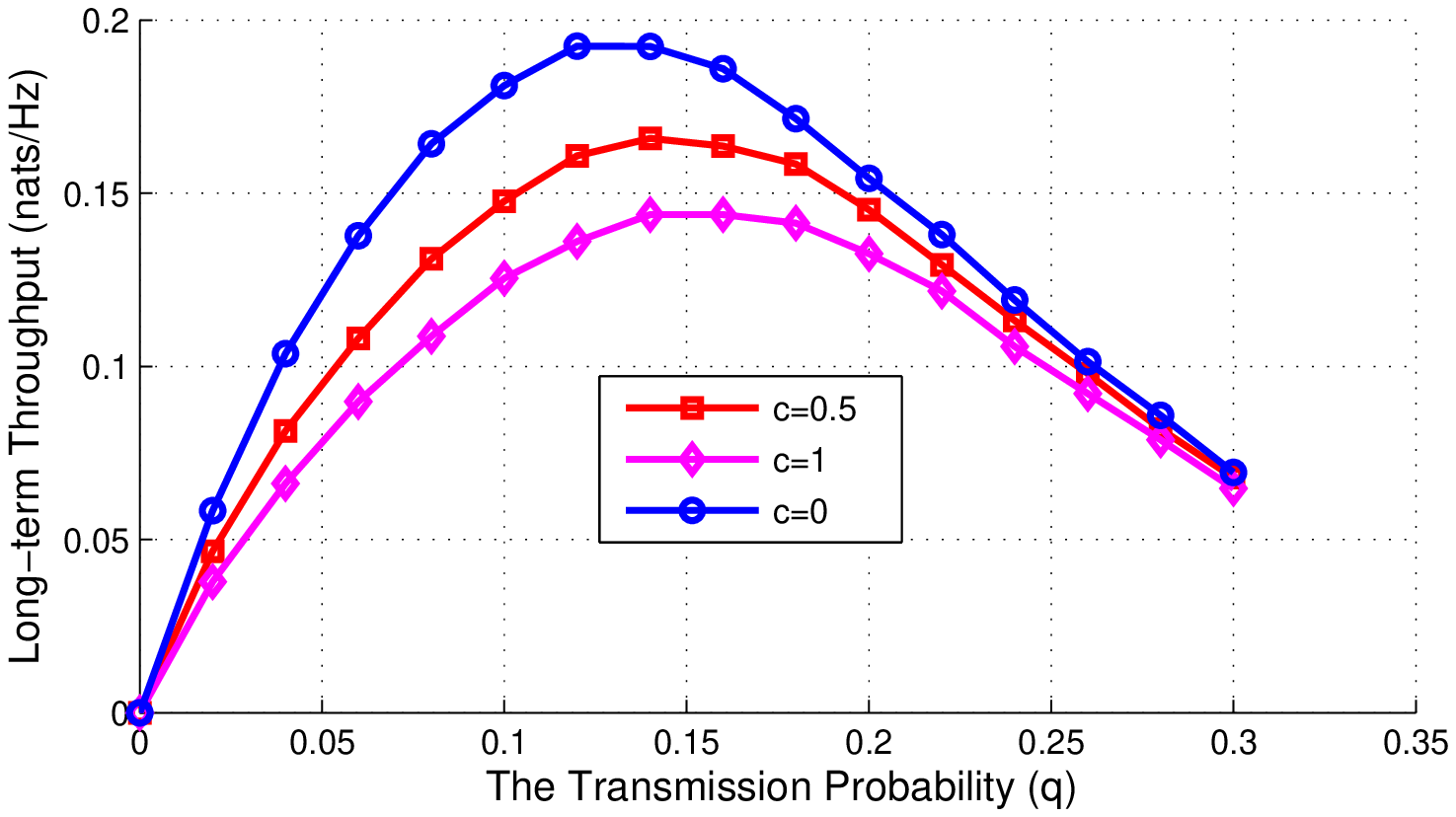}
\label{fig:q}}} \caption{Performance Analysis of the Proposed Algorithm}
\end{figure*}

In our numerical evaluations, we consider a network of ten transmitters. The power
gains are exponentially distributed with mean one for all links, i.e., $\mu_i = 1$ for all $i$. The noise normalized transmit power is taken to be $P=1$.  Furthermore, the portion of time-slot allocated for sensing and the energy used for sensing and pilot transmissions are taken as $\tau = 0.01$\footnote{With 10kHz available bandwidth, the number of samples collected for channel estimation, is 100, which is high enough for perfect estimation.}, and $\phi_{se} = \phi_{pi} = 0.01 \mbox{J/slot}$. The efficiency factor of EH circuitry, $\eta$, is selected as $0.7$. Note that numerical results strongly depend on the system parameters and the structure of $\phi_{de}(\cdot)$. In the following, we focus on a particular energy consumption model, and similar considerations can be made in other cases. Consider $\phi_{de}(\gamma(t)) = c\times \log(1+\gamma(t))+0.5 \mbox{J/slot}$, where $c$ is dependent on the encoding/decoding method utilized in the network and the results are evaluated for different values of $c$. 


In Fig.\ref{fig:V}, we investigate the effect of system parameter $V$
in our dynamic control algorithm for different values of $\phi_{de}$. We take $q_i = 0.1$ for all
nodes, which is the optimal transmission probability without any energy constraint. Total throughput increases with increasing
$V$ and Fig.\ref{fig:V} shows that the long-term throughput for $V > 100$
converges to optimal value fairly closely for all values of $c$ verifying the results of Theorem 1. Furthermore, the long-term throughput decreases with increasing $c$, since the receiver needs to allocate more time-slots for energy-harvesting to decode the same amount of information. Next, we analyze the effect of the transmission probability, $q$, on the long-term throughput in Fig. \ref{fig:q}. Interestingly, transmission probabilities achieving the maximum long-term throughput increases with increasing $c$. This is because, for larger $c$, the receiver needs to harvest more energy to decode the same amount of information, and higher number of transmission instances gives the receiver the opportunity to harvest energy in a larger amount of time instances.

}
\vspace{-0.1in}
\section{Conclusion}
A major contribution of this letter is the introduction of a energy-harvesting receiver in a multi-access channel with collision model. We define our problem as the maximization of the average rate of information, successfully decoded at the receiver. Capitalizing on Lyapunov optimization framework, we propose a dynamic scheme, which establishes an asymptotically optimal solution. Numerical results confirm the efficacy of the proposed scheme. As a future work, we will consider a joint control of transmitters and receivers with the aim of maximization of the network performance with energy consumption constraints. Another direction is to consider energy harvesting with power splitting, where the receiver can perform both information decoding and energy transmitting from the same wireless transmission.

\bibliographystyle{IEEEtran}
\bibliography{d2dcom,macros_abbrev,macros}

\appendices
\vspace{-0.1in}
{\allowdisplaybreaks\vspace{-0.15in}
\section{Proof of Lemma \ref{lemma:prob}}
\label{proof:prob}

In a collision channel model, the event of successful decoding corresponds to the case when only one transmitter accesses the channel. Then, the probability of 
successful decoding in time-slot $t$ can be calculated as:
\vspace{-0.1in}
\begin{align}
P_s(t) &= \Prob{a(t) = 1 | \gamma(t) = x } = \frac{ \Prob{a(t) = 1, \gamma(t) = x }} {\Prob{\gamma(t) =x}}  = \frac{ \sum_{i \in A_k, |A_k| = 1} \Prob{\gamma(t) = x | A_k} \Prob{A_k}} { \sum_{k=1}^{2^N-1} \Prob{\gamma(t) =x | A_k} \Prob{A_k}}, \label{eq:success_prob}
\end{align}
where $a(t)$ is the number of active transmitters in time-slot $t$, and $A_k$ is the $k$th subset of active transmitters. Furthermore, $|A_k|$ denotes the cardinality of subset $A_k$. 

The distribution of the received power given the set of active transmitters in time-slot $t$ follows the distribution of the sum of exponential random variables resulting in a hypo-exponential distribution \cite{Ross07}, i.e.,
\begin{align}
\Prob{\gamma(t) =x | A_k} = \sum_{i \in A_k }\mu_i^{-1} e^{-\frac{x}{\mu_i}}\prod_{j \in A_k} \frac{\mu_i}{\mu_i-\mu_j},
\label{eq:gamma}
\end{align}
 By inserting \eqref{eq:gamma} into \eqref{eq:success_prob}, we obtain the result in the lemma.}
\vspace{-0.1in}
{\allowdisplaybreaks\section{Proof of Theorem \ref{thm:optimalcontrol}}
\label{proof:optimalcontrol}

{\bf Part (1):} It is easy to see that whenever $E(t) \geq \theta$, then the receiver always decides to decode. Hence, assume that $E(t) = \theta - \epsilon$, where $\epsilon$ is a small positive constant. Then, 
\begin{align*}
 E(t+1) &\leq E(t) + (1-\tau)\eta\gamma(t) = \theta - \epsilon  + (1-\tau)\eta\gamma(t) < \theta + \gamma_{max}
\end{align*}
The above inequality is valid for all time-slots. We now show that under our control policy, when $E(t) < \phi_{de}(\gamma_{max}) + \phi_{se}+\phi_{pi}$, then $\rho(t) = 0 $ for all $t$. When $E(t) < \phi_{de}(\gamma_{max})+ \phi_{se}+\phi_{pi}$ and $\theta = \frac{V}{\eta} + \phi_{de}(\gamma_{max}) + \phi_{se}+\phi_{pi}$, the following inequalities hold
\begin{align}
-(E(t) -\theta)((1-\tau)\eta\gamma(t) +\phi_{de}(\gamma(t)))  
& > \frac{V}{\eta} ((1-\tau)\eta\gamma(t) +\phi_{de}(\gamma(t)))  \geq V (1-\tau)\gamma(t) \label{eq:1}\\
& \hspace{-0.2in} \geq V (1-\tau)\log(1+ \gamma(t))  \geq V (1-\tau)\log(1+ \gamma(t))P_s(t). \label{eq:2}
\end{align}
\eqref{eq:1} follows the definition of $\theta$ and $E(t)$, and \eqref{eq:2} follows from the fact that $x \geq \log(1+x)$ for all positive values of $x$. Under our proposed policy, the above inequalities indicate that $\rho(t) = 0$, whenever $E(t) < \phi_{de}(\gamma_{max})+ \phi_{se}+ \phi_{pi}$.

{\bf Part (2):} We first calculate one-step Lyapunov drift as:

\vspace{-0.2in}
\small
\begin{align*}
\Delta(t) &= \frac{1}{2}\E{ (E(t+1)-\theta)^2 -  (E(t)-\theta)^2 | E(t)} = \frac{1}{2} \mathbb{E}\left[ (\rho(t)\phi_{de}(\gamma(t)))^2 + (\phi_{se}+\phi_{pi})^2  ((1-\rho(t))(1-\tau)\eta\gamma(t))^2 \right. \\
& \hspace{1.75in}\left. -2(E(t)-\theta)((1-\rho(t))(1-\tau)\eta\gamma(t) - \rho(t)\phi_{de}(\gamma(t)) -\phi_{se}-\phi_{pi})  | E(t) \right] \\
&\leq B - \E{ (E(t)-\theta)((1-\rho(t))(1-\tau)\eta\gamma(t) -\phi_{se} -\phi_{pi}- \rho(t)\phi_{de}(\gamma(t)) ) | E(t)},
\end{align*}
\normalsize where $B = \frac{\gamma_{max}^2 + (\phi_{de}(\gamma(t)))^2 + (\phi_{se}+\phi_{pi})^2}{2}$. Then, we can rewrite the throughput-mixed Lyapunov drift in \eqref{eq:deltawithreward} as:

\vspace{-0.35in}
\begin{align} \Delta^U(t) & \leq B -  (E(t)-\theta)((1-\rho(t))(1-\tau)\eta\gamma(t) - \rho(t)\phi_{de}(\gamma(t)) ) \nonumber \\
&- V\E{(1-\tau)\log(1+\gamma(t))\rho(t)S(t) | \gamma(t),E(t)}, \label{eq:deltawithreward1}
\end{align}

Our proposed dynamic network control algorithm is designed
such that it minimizes the right hand side of \eqref{eq:deltawithreward1}. Furthermore, Part (1) shows that energy availability constraint in \eqref{eq:energy_const} is naturally satisfied under the proposed algorithm. Then, by using the steps in Theorem 4.5 of \cite{Georgiadis}, one can easily prove Theorem \ref{thm:optimalcontrol}.}

\end{document}